# Harmony in the Australian Domain Space

Xian Gong
xian.gong@student.uts.edu.au
University of Technology Sydney
Sydney, Australia

Paul X. McCarthy
paul@onlinegravity.com
The Data Science Institute
University of New South Wales
Sydney, Australia

Marian-Andrei Rizoiu
Marian-Andrei.Rizoiu@uts.edu.au
University of Technology Sydney
Sydney, Australia

Paolo Boldi
paolo.boldi@unimi.it
Computer Science Department
Università degli Studi di Milano
Milan, Italy

## ABSTRACT

In this paper we use for the first time a systematic approach in the study of harmonic centrality at a Web domain level, and gather a number of significant new findings about the Australian web. In particular, we explore the relationship between economic diversity at the firm level and the structure of the Web within the Australian domain space, using harmonic centrality as the main structural feature. The distribution of harmonic centrality values is analyzed over time, and we find that the distributions exhibit a consistent pattern across the different years. The observed distribution is well captured by a partition of the domain space into six clusters; the temporal movement of domain names across these six positions yields insights into the Australian Domain Space and exhibits correlations with other non-structural characteristics. From a more global perspective, we find a significant correlation between the median harmonic centrality of all domains in each OECD country and one measure of global trust, the WJP Rule of Law Index. Further investigation demonstrates that 35 countries in OECD share similar harmonic centrality distributions. The observed homogeneity in distribution presents a compelling avenue for exploration, potentially unveiling critical corporate, regional, or national insights.

## 1 INTRODUCTION

The broad field of computational social science aims to use collected data of various kinds (most notably, online data) to explore different aspects of economics, history, and psychology. The type of analysis that can be performed on the collected data may have different purposes and the tools used are also varied, ranging from simple statistics to complex machine learning models. The Web is a particularly interesting source of data, as it is a large-scale, real-world, and continuously evolving system. Data about the Web can be collected in various ways, including crawling, scraping, and mining. There are many public collections of web crawls, but one that is known for being very reliable and quite wide in scope is the Common Crawl[1]. Common Crawl's measurements are preferred for web and network analysis due to their extensive coverage, regular updates, and large-scale, publicly accessible datasets, which reduces the need for resource-intensive data collection and is applicable across various research in a reproducible way.

One important and extremely relevant phenomenon that social science is considering recently is the decline in economic diversity. Economic diversity refers to the variety and distribution of different types of economic activities and industries within a region or economy, reflecting its breadth and resilience against market fluctuations. Research shows that more diversified economies are associated with higher levels of gross domestic product, lower unemployment rates and less instability [13, 22].

The exploration of web graph data is emerging as a powerful tool for investigating economic diversity. A large-scale study that looked at over 10 years of web data by McCarthy et al. [15] reveals a clear trend: a steady long-term decrease in global link diversity. This decline is attributed to the growing market hegemony of a few firms, as captured by the Herfindahl-Hirschman Index. This trend signifies an increasing monopolization of the market by a small number of corporations, raising critical questions about the future of economic diversity.

In addition to its value for macroeconomic analysis, web graph data at a domain or page level proves instrumental in investigating firm-level trends, with various measures of network centrality serving as essential tools. Comprehending the patterns in the centrality of domains within the web graph not only facilitates firm-level analysis but also potentially elucidates broader trends across the economy.

In Web analysis, various measures of centrality can be used (both at a page or at a domain level of granularity), with PageRank [19] having a major role; nonetheless, recently other centrality measurements, and especially harmonic centrality [3, 5, 18], were suggested as crucial for identifying influential web pages and understanding their role in the information flow and structure of the internet network. Previous research [19] already showed that harmonic centrality can be used to analyze economic networks, identifying key interconnected industries or regions, thereby providing insights into the structure and resilience of an economy's diversity [10].

---
[1]https://commoncrawl.org/

In this paper, we use for the first time a systematic approach in the study of harmonic centrality at a Web domain level, and gather a number of significant new findings. In particular, we explore the relationship between economic diversity and the structure of the Web within the Australian domain space, using harmonic centrality as the main structural feature. From a more global perspective, harmonic centrality can be used at a country level to look at the relative global centrality of all domains within a top-level country domain space [16] and we find a significant correlation (coefficient = 0.45 at a 5% significance level) between the median harmonic centrality of all domains in each OECD country and one measure of global trust, the WJP Rule of Law Index. Further investigation demonstrates that 35 countries in OECD share similar harmonic centrality distributions (Figure 1). The observed homogeneity in distribution presents a compelling avenue for exploration, potentially unveiling critical corporate, regional, or national insights.

Our focus here is mainly on the relationship between firm-level characteristics and network structure within the Australian domain space. At the beginning of our journey, we were interested in whether the multi-peaked distribution of harmonic centrality observed in the most recent snapshot of the Australian web space held true at different time points, which turned out to be true. The next step was decomposing the entire harmonic centrality distribution and analyzing the quantitative and qualitative characteristics between different peaks. Finally, we use statistical analysis and machine learning algorithms to explore whether the harmonic centrality measurements contain information not explicitly stated in the web graph, such as the company's revenue, expenses, and social media links. These non-structural features are closely tied to its economic diversity, influencing financial stability, cost management, and market reach.

In our analysis, we found that harmonic centrality scores are complex measures that combine information related to the network graph structure, such as the size of domains and the number of inbound links, but also include information related to offline unstructured characteristics such as the size, scope, influence, and impact of the organization. The contribution of this paper is to explore the nature of harmonic centrality further, examine structural and nonstructural features associated with these measurements, find significant correlations, and provide another view to understanding harmonic centrality. We illustrate that it not only measures the centrality of a node in a network, providing insight into its influence and connectivity within the network but also contains other implicit information in the web graph, which can be used to classify a company and predict its economic performance. Its interconnectedness and influence within a network can reflect its economic diversity, as companies with a wide range of products or services often have diverse market connections and are more resilient to market changes. This centrality can also highlight opportunities for strategic partnerships and market expansion.

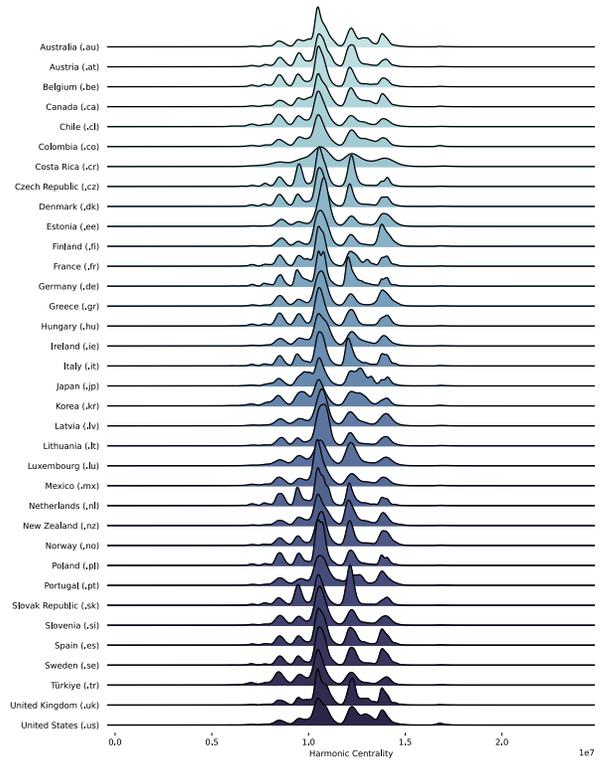

**Figure 1: Empirical Harmonic Centrality Distributions of Web Domains among 35 Countries in 2023.**

## 2 BACKGROUND AND RELATED WORK
### 2.1 Harmonic Centrality

Harmonic centrality [2] is a measure of centrality of nodes in a graph which was proposed to solve the issue of unreachable vertices in closeness centrality [5]. In particular, we can define the *harmonic centrality* of a node $u$ in a directed graph $G$ as

$$h(u) = \sum_{v \in N_G \setminus \{u\}} \frac{1}{d_{uv}},$$

where $N_G$ is the set of nodes of $G$, $d_{uv}$ is the length of the shortest path from $u$ to $v$ and we assume $1/\infty = 0$. Harmonic centrality has a number of theoretical and practical qualities (see, for instance, [4, 5]) that make it quite promising as a measure of importance in general directed networks.

How harmonic centrality values are distributed in certain real-world networks is a underexplored area of research. Since we are using harmonic centrality distribution within domain graphs as a way to measure some underlying phenomena, we wonder what is the *natural* distribution we can expect.

In general, understanding the reasons behind (or even just reproducing) certain observed behaviors and distributions of some characteristics of real graphs is challenging, even in the case of very simple features such as in- and out-degrees—this is, in fact, the main quest of the theory of random graph models of complex networks [1, 23].

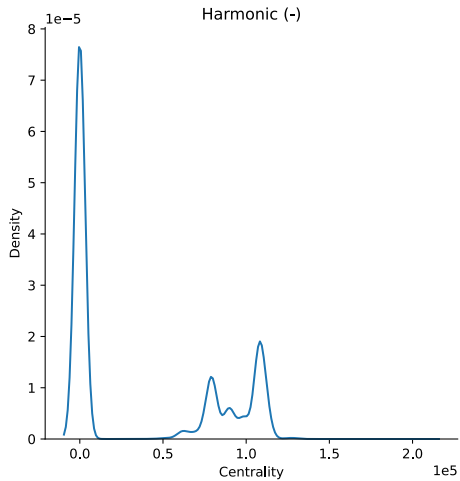

**Figure 2: Expected distribution of harmonic centrality values according to a theoretical model and a sample network graph with** 700 000 **nodes. [6]**

A well-known scale-free model of directed networks which is expected to represent faithfully web graphs and their offspring is described in [6]: it is an evolving network model, where the network grows one node/arc at every step, based on four parameters called $\alpha$, $\beta$, $\delta^-$ and $\delta^+$: with probability $\alpha$, a new node $v$ is added along with an arc *to* an existing node, chosen based on its in-degree; with probability $\beta$, a new arc *between* two existing nodes is added, whose source and target are chosen based on their out- and in-degree, respectively; with probability $1 - (\alpha + \beta)$, a new node $v$ is added with an arc coming *from* an existing node, chosen based on its out-degree. When choosing a node based on its in- and out-degrees we in fact adopt a bias, and $\delta^{\pm}$ represent the two biases (see [6] for details).

Now, using the same analysis as in Section 4 of [6], we assume that $\delta^+ = 0$ (i.e., that outdegrees are unbiased), obtaining the following relations between in-degree exponent $X^-$, outdegree exponent $X^+$ and the other parameters:

$$\alpha = \frac{X^+ - 2}{X^+ - 1}$$
$$\alpha = \frac{1 + \delta^-(1 - \beta)}{\alpha + \beta}.$$

Based on the measurements of [17], in- and out-degrees of host graphs fit well a power-law distribution with in-degree exponent $X^- = 2.12$ and out-degree exponent $X^+ = 2.14$, which gives

$$\alpha \approx 0.12281$$

and values of $\beta$ in the range $[0.77, 0.87]$. Taking $\beta$ in the midpoint of the range (i.e., $\beta = 0.82$), we obtain $\delta^- = 0.31078$. For these values of the parameters, the distribution of harmonic centrality values is shown in Figure 2 (on a sample with 700 000 nodes). The large mode on 0 is an artifact of the model, which produces many isolated nodes. However, the distribution corresponds reasonably well to the observed multi-modality, although in our datasets there are more modes and the corresponding densities are more similar to one another.

## 2.2 Applications on Firm-Level

In this study, we apply the concept of harmonic centrality at the domain (website) level, with the aim of providing valuable insights into the structure and dynamics of the web at a firm level. It can be a useful tool for understanding firm-level economic diversity.

One previous study used harmonic centrality to analyze structures of other possible networks, such as global supply chains [12]. Here researchers mined data from financial records to construct the global supply chain network of the auto manufacturing sector and explored the effect of these supply-chain centrality measures on firms' financial performance, investment risk, and market value volatility.

In another study [21] authors introduced a new normalized firm centrality measure independent of corporate group size. This centrality notion was computed on a set of global corporate networks encompassing 17.8 million firms. They found a positive relationship between firm centrality and firm performance, but the significance of the relationship decreases as corporate group size increases.

Recent research, grounded in the concept of harmonic centrality, introduces two proximate metrics to examine the connection between senior executives' social connections and the firm's value [20]. This study demonstrates the significant impact of top executives' social networks on the firm's market capitalization by correlating the Chief Marketing Officer's network centrality with the company's market value.

While various centrality metrics are employed to scrutinize distinct networks, the underlying objective remains consistent. The aim is to unearth effective measures for evaluating firm-level data to assist in decision-making processes. Though not focused on localized networks, the extensive and all-encompassing web graph from Common Crawl utilized in our study enables the extraction of implicit information on a broader scale. For businesses, harmonic centrality stands out as a more comprehensible and readily accessible metric.

## 3 DATA AND EXPERIMENTS

This section describes the sources of datasets used for this study. We describe the structural and non-structural features used for the experiments in Table 1 (Section 3.1), the design of the clustering analysis (Section 3.2), and the statistical analysis (Section 3.3) followed by classification prediction (Section 3.4).

## 3.1 Dataset Description

We utilize the datasets mainly from Common Crawl and BuiltWith in our study. Common Crawl provides the domain-level graph that encapsulates the relationships among over 88 million domains (websites) found on the World Wide Web across different periods. It is a network representation of domain interconnections, offering insights into the web's link structure, domain attributes and the prominence of websites within the web ecosystem. Harmonic centrality is one of the structural features of the domain-level graph. As explained above, it is based on the reciprocal distance between nodes, emphasizing the importance of nodes that are closer to other nodes in terms of network traversal. Moreover, it excels at identifying nodes with strong local influence within tightly-knit communities or clusters. Unlike PageRank, it is less susceptible to manipulation and provides an intuitive interpretation, measuring the average reciprocal distance between a node and all others, which is particularly useful for detecting community structures and analyzing dynamic networks.

Table 1: Descriptions of Structural and Non-Structural Features of Australian Domains

|  | Features | Description |
| --- | --- | --- |
| Structural | Harmonic Centrality | Node importance in a network depends on close connections, especially shorter distances. |
|  | Domain Size | The number of web pages or websites hosted under that specific domain name. |
|  | inbound | The number of incoming links or connections to a specific web page or website. |
|  | outbound | The number of links or connections originating from a specific web page or website. |
|  | Tranco | Website popularity and ranking determined by web traffic and domain authority assessment. |
| Non-Structural | Tech Spend | Monthly expenditure estimate on tech services and tools for websites or online businesses. |
|  | Sales Revenue | An estimate of the monthly or annual sales revenue of eCommerce websites. |
|  | Social | Estimation of social media links, encompassing Facebook, Twitter, Google+, Pinterest, GitHub, and LinkedIn. |
|  | Employees | An estimate of amount of employees for websites. |
|  | Age | Years since this website was first indexed until now. |

BuiltWith is a web technology profiling and analysis tool that focuses on providing insights into the technological infrastructure of websites. Its core functionality involves detecting and reporting on the web technologies, frameworks, content management systems, and software tools utilized by websites. BuiltWith offers users the ability to ascertain the structural elements of web development, including programming languages, web servers, and security measures. While it primarily addresses the structural aspects, it provides basic information about non-structural features of domains, such as technology spending, sales revenues, and employees, which offer insights into domains and economic indicators.

Common Crawl and BuiltWith have different focuses and capture different types of data, and their approaches and frequencies to tracking dynamic changes of features over time vary. Common Crawl primarily focuses on capturing the static content of web pages and typically conducts large-scale crawls every few months to update its datasets. BuiltWith specializes in profiling the web technologies and software tools used by websites. It does not capture dynamic changes in website content or features over time. Instead, it provides a snapshot of the technologies in use at the time of analysis. Even though there is a time mismatch between Common Crawl and BuiltWith, we assume that the features of domains will remain stable for a short period. Therefore, our further analysis uses the most recent domain ranks from Common Crawl and the current snapshot of basic information from BuiltWith.

We obtained the domain name and its Harmonic Centrality from the Common Crawl website for the last quarter of each of the six years from 2018-2023. Since this article only focuses on the Australian domain name space, we filter the entire Common Crawl websites by domain names ending in .au. The number of Australian domain names in these six years ranged from 744k to 977k. The number of domain names in Australia remained above 900k in the first five years but will drop to 744k in 2023. The sharp decline in Australian domain names between 2022 and 2023 is primarily due to the introduction of .au direct domains, which allows entities to register shorter .au domain names. This resulted in a restructuring of domain registration in Australia, probably reducing the use of .com.au, .net.au, and .org.au domains as people moved to the shorter .au domain. We use Australian domains in 2023 as a benchmark to obtain features from BuildWith due to its snapshot characters. We found 744,058 Australian domain names in 2023, and based on this we collected the features of Table 1.

## 3.2 Clusters Movements over time

In our investigation of the Australian domain space, we conducted an extensive analysis of harmonic centrality data obtained from the Common Crawl dataset spanning the years 2018 to 2023. Our examination primarily focuses on the harmonic centrality distributions observed in the first quarter of each of these six years. As illustrated in Figure 3, it is evident that the harmonic centrality distributions exhibit a consistent pattern across the different years. To gain deeper insights into the distinctive peaks within these distributions, we employed a clustering algorithm, specifically KMeans, to decompose each distribution. Using clustering algorithms to analyze data with multimodal distributions over time is beneficial because these algorithms can effectively identify and separate distinct patterns or trends within the data. This approach facilitates a deeper understanding of the dynamics and variability in the data over time, which is essential for concretely observing the changes in domains over the past six years and analysing the changes in those non-structural features under different movement scenarios. To determine the optimal number of clusters for each year, we utilized the average of the Davies–Bouldin index and the Calinski–Harabasz index. The Calinski-Harabasz index [7] assesses the ratio of within-cluster dispersion to between-cluster dispersion for all clusters, whereas the Davies-Bouldin index [9] quantifies the similarity between clusters, thereby allowing us to evaluate the separability of categories. To align the Davies–Bouldin index with the Calinski–Harabasz index for ease of interpretation, we took the reciprocal of the former, with a higher value indicating superior clustering performance, and then we combined the two indices by taking their average.

In Figure 4, the blue line represents the average combined index along these six years, while the grey area denotes the one-standard deviation interval from the mean. Notably, the score reaches its maximum when selecting six clusters, after which it gradually diminishes. Consequently, we opted to employ six clusters to partition the annual distributions, with the cluster possessing the lowest harmonic centrality value designated as Cluster 0, and the cluster with the highest harmonic centrality value labelled as Cluster 5, following an ascending order. This arrangement results in six clusters with fixed relative positions, and the harmonic centrality values ascend sequentially. With each domain name assigned to its respective clustering category for each year, we were able to analyze the temporal movement of domain names across these six positions. This analysis aims to discern whether such movements yield insights into the Australian Domain Space or exhibit correlations with other characteristics.

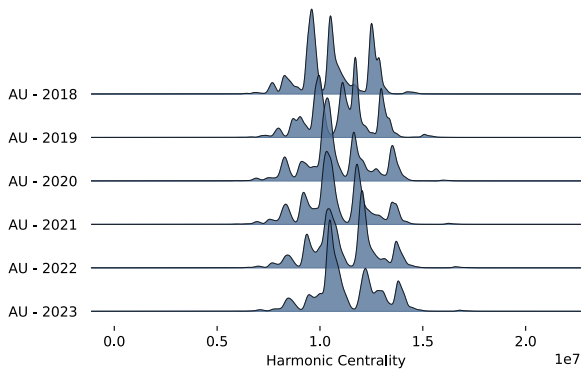

Figure 3: Harmonic Centrality Distribution Over Time from 2018 to 2023 within the Australia (.au) Web Domain.

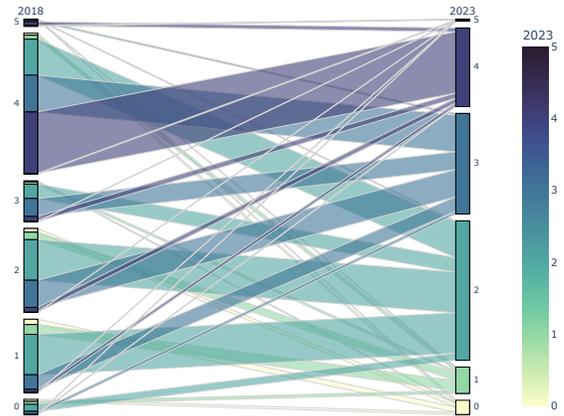

Figure 5: Cluster Movements of Australian Domains between 2018 and 2023.

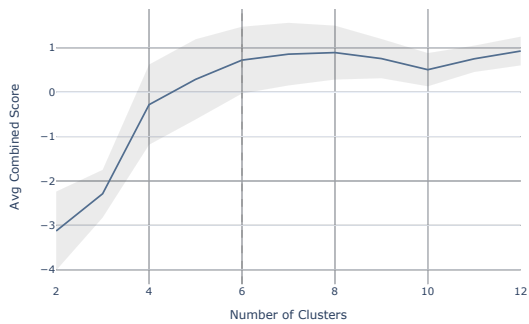

Figure 4: Six clusters appear as the optimal number within the Harmonic Centrality Distributions. The graph shows the average and standard deviation of combined scores (Calinski-Harabasz and inverse Davies-Bouldin indices) on KMeans performance on harmonic centrality distributions of Australian Web Domain Space from 2018 to 2023.

For instance in Figure 5, when scrutinizing cluster movements at the two extremes of the Australian domain space timeline between 2018 and 2023, we observed that the sources of most clusters were dispersed with the exception of Cluster 4 (the majority of current domain names in Cluster 4 originated from Cluster 4 six years prior). Additionally, the sizes of high harmonic centrality clusters in 2018 (such as Clusters 4 and 5) were notably larger than their counterparts in 2023, suggesting a trend towards greater concentration of the Australian web space on a selected few domain names. For a more comprehensive statistical summary of our findings, please refer to Section 4.

### 3.3 Statistical Analysis within Clusters

This section presents our process for conducting statistical analyses of harmonic centrality in conjunction with the structural and non-structural attributes outlined in Table 1. Our primary objective is to investigate a statistically significant linear association between harmonic centrality and these attributes within the six clusters. We employ three distinct statistical tests to evaluate these hypotheses: Pearson correlation, Spearman correlation, and Kendall rank correlation. Pearson Correlation is a statistical metric designed to quantify the strength and direction of the linear relationship between two continuous variables. On the other hand, Spearman Correlation is a non-parametric statistical approach that assesses the magnitude and direction of the monotonic connection between two variables. Kendall Rank Correlation, also referred to as Kendall's Tau, is another non-parametric statistical technique employed to gauge the strength and direction of the association between two variables. These three distinct statistical measures are applied in tandem to analyse the relationships between variables comprehensively. Pearson correlation proves effective in capturing linear relationships, while Spearman and Kendall's correlations are particularly valuable for detecting monotonic relationships, including those that may be non-linear in nature. We broaden our scope by employing all three measures to encompass a wider array of potential connections between variables. This approach enhances the robustness and informativeness of our analyses, especially in scenarios where the nature of the relationship is uncertain or when the data do not adhere to assumptions of normal distribution.

To acquire representative samples from the six sub-distributions for the purpose of hypothesis testing, we opted to select data falling within one standard deviation of the mean as our sample set. It is worth noting that any instances with missing values were excluded from consideration during the testing of individual features. Consequently, the sample sizes for each feature may not be uniform, but we took measures to ensure that the sample sizes remained sufficiently large for statistical analysis.

### 3.4 Clusters Prediction

Our primary focus lies in examining the significance of non-structural attributes in determining the classification of Australian domains into specific clusters. These non-structural attributes contain information not encompassed within the web graph and are intricately linked to a company's economic performance. Furthermore, our statistical analysis has revealed that these non-structural attributes exhibit a non-linear association with harmonic centrality values.

In this segment of our experimentation, we have employed a machine learning classification algorithm to assess the predictive capabilities of these non-structural attributes and establish their relative importance. Specifically, we have utilized the XGBoost classifier and fine-tuned its parameters for this purpose. The XGBoost Classifier [8] is renowned for its exceptional accuracy and efficiency in machine learning tasks. It leverages decision trees to address intricate data relationships, facilitates feature selection, and mitigates overfitting, rendering it a versatile choice for handling classification tasks. Although SVM and Random Forest allow better explainability, XGBoost offers advantages over them in its ability to efficiently handle large and sparse datasets, offer higher performance with features like regularization and custom optimization, and generally deliver better accuracy, especially for complex problems. However, the effectiveness of these algorithms can vary based on the specific nature of the data and the problem at hand. In subsequent research, we can introduce other classifiers and compare the performances.

It is evident that the categories within our dataset exhibit a pronounced imbalance. Initially, the classifier is trained to utilize the original dataset. Subsequently, a method of random sampling is employed to extract 2,000 samples from each cluster, thereby facilitating the creation of a new, balanced dataset. This newly constituted dataset is then utilized to retrain the classifier. A comparative analysis is then conducted to evaluate the variations in accuracy between the classifiers trained on the original unbalanced dataset and the newly balanced one.

## 4 RESULTS

In this section, we consolidate the empirical findings from Section 3 into a coherent analytical narrative, encompassing the quantitative and qualitative dimensions.

### 4.1 Quantitative Findings

*Basic characteristics of six clusters.* Our experimental analysis dissected nearly six years of harmonic centrality distributions into six distinct sub-distributions defined by specific clustering criteria. This division categorizes the trajectory of domain movements over the six periods into three scenarios: firstly, domains remaining static within their initial cluster (constant); secondly, domains ascending from clusters of lower harmonic centrality to those with higher values (upward moving); and thirdly, domains descending through the cluster hierarchy (downward moving).

Table 2 details the mean values or ratios of non-structural features across varied scenarios. The 'Current cluster' in the table presents the count of domain samples per cluster for 2023, along with the average value of each feature. The 'Constant cluster' contains domains of six clusters that have stayed within the same cluster across the six periods. This row still demonstrates the number of samples in each cluster and the corresponding mean values of features.

The latter two scenarios, 'Upward moving' and 'Downward moving', provide a comparative analysis, examining the mean value ratios between 'moving clusters' and 'Constant clusters'. For example, in the 'Upward moving' scenario, there are 331 domains moved from lower-ranked clusters (# 0 - # 4) to # 5 while there are 424 domains kept staying at the highest-ranked cluster # 5 during the six years. A ratio of 0.22 in the Sales Revenue column of upward moving scenario means that the mean value of sales revenue of the 331 upward-moving domains is divided by the mean value of 424 domains in constant cluster # 5. By analogy, the ratio of 1.19 in the same column means that the mean value of 3,518 domains moved from lower-ranked clusters (# 0 - # 3) to # 4 divided by the mean value of 2,279 domains in constant cluster # 4. The 'Downward moving' scenarios work oppositely. It measures the ratios of down-moving domains to the corresponding low-ranked cluster. For instance, the ratio of 2.96 in the Sales Revenue means that the mean value of 1,077 domains moved from the higher-ranked cluster (only # 5 in this case) to # 4 divided by the mean value of 2,279 domains in constant cluster # 4.

Based on the first two scenarios, domains in a higher-ranked cluster exhibit a corresponding increase in average sales revenue and technological expenditure, following an exponential trend. The 2023 data set was employed to highlight the characteristics of these six distributed non-structural attributes. Figure 6 graphically represents the average values of each non-structural attribute for domains that maintain their position within the same clusters throughout the observed period. This visualization demonstrates an exponential increase in these mean values, concomitant with rising harmonic centrality, particularly pronounced within the uppermost clusters (those with the highest harmonic centrality). Such findings underscore the markedly distinct harmonic centrality levels across the six clusters, hinting at different economic performances at the firm level.

Given the limitations of non-structural feature data (only available in 2023), our analytical scope regarding cluster movement scenarios is inherently constrained. Observations derived from these ratios indicate that with the exception of advancements from the lowest-ranking cluster to the uppermost cluster (#5) — which do not exhibit an enhancement in features exceeding a factor of 1 — ascensions through other clusters manifest a discernible impact on feature mean values. Nevertheless, it is noteworthy that the ratio associated with the declining scenarios generally surpasses that observed in ascending scenarios, suggesting a predominant influence exerted by the directionality of cluster movement.

Table 2: Statistics Summary (Mean/Ratio) of Three Scenarios for Six Clusters

| Scenarios | Cluster (Sample Size) | Sales Revenue (Mean) | Tech Spend (Mean) | Social (Mean) | Employees (Mean) |
|---|---|---|---|---|---|
| Current | # 5 (n=2,965) | 48,323.85 | 1,226.92 | 18,864 | 300 |
| | # 4 (n=108,084) | 8,943.13 | 272.30 | 5,539 | 24 |
| | # 3 (n=187,389) | 5,067.65 | 181.38 | 1,952 | 6 |
| | # 2 (n=310,480) | 3,008.84 | 101.64 | 1,169 | 1 |
| | # 1 (n=81,761) | 1,909.26 | 78.93 | 303 | 1 |
| | # 0 (n=53,379) | 2,121.93 | 74.33 | 616 | 0 |
| Constant | # 5 (n=424) | 120,108.77 | 3,188.61 | 48,737 | 1,153 |
| | # 4 (n=2,279) | 10,622.74 | 344.03 | 3,735 | 8 |
| | # 3 (n=3,207) | 4,265.89 | 203.16 | 791 | 8 |
| | # 2 (n=4,011) | 2,006.97 | 66.52 | 457 | 0 |
| | # 1 (n=502) | 1,026.21 | 32.95 | 162 | 0 |
| | # 0 (n=230) | 0.00 | 7.50 | 0 | 0 |

| Scenarios | Cluster (Sample Size) | Sales Revenue (Ratio) | Tech Spend (Ratio) | Social (Ratio) | Employees (Ratio) |
|---|---|---|---|---|---|
| Upward Moving | to # 5 (n=331) | 0.22 | 0.29 | 0.07 | 0.04 |
| | to # 4 (n=3,518) | 1.19 | 1.17 | 1.07 | 0.61 |
| | to # 3 (n=7,540) | 2.51 | 1.61 | 2.56 | 0.90 |
| | to # 2 (n=2,851) | 4.24 | 3.51 | 1.80 | 2.10 |
| | to # 1 (n=240) | 3.22 | 4.36 | 0.43 | |
| Downward Moving | to # 4 (n=1,077) | 2.96 | 2.87 | 1.91 | 10.57 |
| | to # 3 (n=3,624) | 3.02 | 1.94 | 3.42 | 0.99 |
| | to # 2 (n=4,406) | 3.66 | 3.49 | 12.59 | 8.05 |
| | to # 1 (n=1,457) | 3.31 | 3.61 | 2.74 | |
| | to # 0 (n=420) | | 7.51 | | |

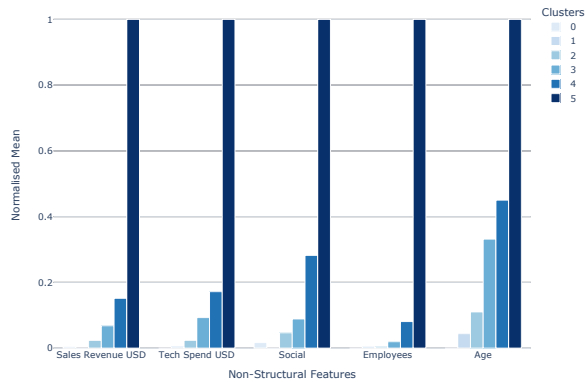

Figure 6: Significant different means of non-structural features among six clusters of AU Domain Space in 2023.

*Correlations between harmonic centrality and (Structural and Non-structural) Structural features.* In addition to comparing the mean values among six clusters, we also explored the linear relationship between harmonic centrality and structural and non-structural features under each cluster. We summarise the results in Table 3 3. In clusters exhibiting large harmonic centrality values, such as clusters 3, 4, and 5, the majority of hypothesis tests display statistical significance. However, numerous missing values within the three characteristics of Tranco, Social, and Employees introduce a degree of uncertainty, potentially making some of these tests non-statistically significant. This observation suggests a need for a cautious interpretation of these results, acknowledging the limitations imposed by the incompleteness of the data for these specific attributes. Interestingly, clusters with low harmonic centrality values, such as clusters 0 and 1, negatively correlate significantly with Sales Revenue. Observations indicate that for certain features, including inbound links, Tech Spend, Sales Revenue, and Social metrics, the Spearman and Kendall rank correlation coefficients exhibit higher values than the Pearson correlation coefficient. This discrepancy suggests the existence of a robust non-linear relationship between these variables and harmonic centrality, emphasizing the complexity of their interdependencies beyond linear associations.

*Insights from the classification prediction.* Using the XGBoost classification algorithm with tuning and cross-validation, we studied the predictive efficacy of non-structural features in six clusters since we are more interested in the relationship between the harmonic centrality of structural features and those non-structural features. The f1 score of the classifier trained on the original dataset is 0.22, while the f1 score of the classifier trained on the balanced dataset is 0.29. The accuracy has not improved much, but we have very low accuracy for clusters 0 and 5 in the first case. In the second case, the prediction accuracy for clusters 0 and 5 is much improved. Figure 7 demonstrates the importance of the feature as ranked by XGBoost. The above two classifiers have the same order of importance for non-structural features. The sparsity of our feature matrix, coupled with the exclusive reliance on non-structural features for model training, has resulted in suboptimal accuracy levels. However, an analysis of feature importance reveals that company information, not encapsulated within the web graph, does exert a discernible influence on the predictive capability regarding the classification of domain names into categories of high or low harmonic centrality. This observation underscores the potential impact of external data sources in enhancing predictive accuracy. The demonstrated predictive capability affords a novel vantage point in evaluating a corporation's economic efficacy. Overlooked structural attributes, such as harmonic centrality, emerge as newly identified determinants influencing corporate performance metrics.

Table 3: Correlation Analysis of Structural and Non-Structural Features with Harmonic Centrality. Pearson Correlation (P), Spearman Correlation (S) and Kendall's Tau (K) are used to measure the coefficients. Stars are used to indicate the significance level: 5% (*), 1% (**) and 0.1% (***).

|  | Cluster 0 | Cluster 1 | Cluster 2 | Cluster 3 | Cluster 4 | Cluster 5 |
|---|---|---|---|---|---|---|
| DomainSize (P) |  |  | 0.02 *** | 0.02 *** | 0.01 * | 0.15 *** |
| DomainSize (S) | 0.01 ** |  | 0.05 *** | 0.03 *** | 0.12 *** | 0.52 *** |
| DomainSize (K) | 0.01 ** |  | 0.04 *** | 0.03 *** | 0.09 *** | 0.4 *** |
| inbound (P) | 0.04 *** | 0.05 *** | -0.02 *** | 0.03 *** | 0.11 *** | 0.52 *** |
| inbound (S) | 0.06 *** | 0.05 *** | -0.09 *** | 0.15 *** | 0.38 *** | 0.84 *** |
| inbound (K) | 0.05 *** | 0.04 *** | -0.05 *** | 0.1 *** | 0.29 *** | 0.69 *** |
| outbound (P) |  |  | 0.04 *** | 0.05 *** | 0.04 *** | 0.2 *** |
| outbound (S) | 0.07 ** |  | 0.1 *** | 0.04 *** | 0.1 *** | 0.39 *** |
| outbound (K) | 0.05 ** |  | 0.07 *** | 0.03 *** | 0.07 *** | 0.27 *** |
| Tranco (P) |  | -0.33 ** |  |  | -0.17 *** | -0.33 *** |
| Tranco (S) |  | -0.34 ** | 0.08 * | -0.07 * | -0.16 *** | -0.38 *** |
| Tranco (K) |  | -0.23 ** | 0.06 * | -0.05 * | -0.11 *** | -0.26 *** |
| Tech Spend (P) |  |  | 0.03 *** | 0.05 *** | 0.05 *** | 0.27 *** |
| Tech Spend (S) |  | 0.01 * | 0.03 *** | 0.06 *** | 0.15 *** | 0.31 *** |
| Tech Spend (K) |  | 0.01 * | 0.02 *** | 0.05 *** | 0.11 *** | 0.23 *** |
| Sales Revenue (P) |  |  | 0.02 *** | 0.02 *** | 0.05 *** | 0.16 *** |
| Sales Revenue (S) | -0.03 * | -0.02 ** | 0.03 *** | 0.02 *** | 0.09 *** | 0.05 * |
| Sales Revenue (K) | -0.02 * | -0.01 ** | 0.02 *** | 0.02 *** | 0.07 *** | 0.04 * |
| Social (P) |  |  | 0.01 ** |  |  | 0.1 *** |
| Social (S) |  |  | 0.06 *** | 0.02 *** | 0.11 *** | 0.26 *** |
| Social (K) |  |  | 0.05 *** | 0.01 *** | 0.08 *** | 0.2 *** |
| Employees (P) |  |  |  |  |  | 0.13 *** |
| Employees (S) |  |  | 0.02 *** | 0.03 *** | 0.02 *** | 0.1 *** |
| Employees (K) |  |  | 0.02 *** | 0.03 *** | 0.02 *** | 0.08 *** |
| Age (P) |  | 0.03 *** | -0.01 *** | 0.04 *** | 0.13 *** | 0.45 *** |
| Age (S) |  | 0.03 *** | -0.02 *** | 0.05 *** | 0.2 *** | 0.48 *** |
| Age (K) |  | 0.02 *** | -0.01 *** | 0.03 *** | 0.14 *** | 0.34 *** |

### 4.2 Qualitative Findings

In this section, we delve into a qualitative examination of each cluster's inherent characteristics, drawing upon specific domain examples that typify the respective clusters[2]. Our focus is particularly on domains that have maintained a consistent position within clusters across all periods, as this consistency aligns with one of the previously identified scenarios in our study. To elucidate these findings, we comprehensively summarise their characteristics and prominent domain company instances in Table 4.

Domains that have steadfastly remained within the highest harmonic centrality clusters for the whole duration of six years are generally well-established, longstanding organizations. These entities are characterized by their iconic and high-profile branding, a testament to their enduring presence and recognition in their respective fields. Furthermore, these organizations typically command high-traffic websites, which serve as digital hubs attracting large, diverse audiences. This aspect reflects their extensive reach and influence, both nationally and globally. Often, these enterprises are headquartered in major metropolitan cities, which serves as a nexus for economic, cultural, and technological advancements, further reinforcing their prominence and impact.

---
[2]The interested reader can get more information from the full dataset, anonymously available at this link.

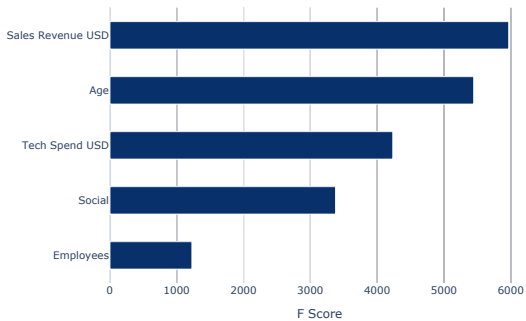

**Figure 7: Feature importance plot of XGBoost Classifier trained on the balanced dataset. The F-score quantifies the relative significance of each feature within the classifier based on how frequently each feature is used to split the data across all trees in the ensemble.**

Conversely, a contrasting picture emerges when examining domains that have consistently been a part of the lowest harmonic centrality clusters. These domains are commonly associated with younger, newly established enterprises. Their genesis in more recent times often means they are still in the nascent stages of brand development and market penetration. These entities tend to be geographically situated in regional or remote areas, focusing on localized markets and specific niches. The nature of their business often results in smaller-scale firms mirrored in their digital presence through low-traffic websites that cater to a more limited, often specialized audience. This dichotomy between the high and low harmonic centrality clusters underscores the diverse spectrum of enterprises and their varying stages of growth, market presence, and digital influence, revealing the multifaceted nature of the economic and digital landscape.

In the second scenario, we focus on domains ascending within the harmonic centrality cluster from 2018-2023. Notable instances of this trend include the Australian pharmacy conglomerate Chemist Warehouse, which, according to LinkedIn data, has been expanding at a rate exceeding 8% annually, now boasting a network of over 550 pharmacies. Similarly, Perfect Portal, the UK-based online legal software provider for small practices, has shown a remarkable annual growth of 18% and extended its operations to Australia. Another example is the men's health charity "Mr Perfect", which adopts a community-oriented network model, orchestrating fundraising events like barbeques. Additionally, the global parking application ZipBy exemplifies this upward trajectory with an impressive 67% annual growth rate. These cases are emblematic of organizations typically experiencing rapid expansion in terms of employee count and sales revenue. They often represent young, globally oriented enterprises venturing into new markets, utilizing models based on franchising, networking, or app-based services, such as mobile parking applications. This pattern highlights a dynamic where organizations leverage digital platforms and global connectivity to foster significant growth and market penetration.

Our analysis of the final scenario focused on domains that exhibited a downward movement in harmonic centrality clusters during 2018-2023. This category predominantly includes government entities undergoing rebranding, restructuring, or dissolution. The Australian Government's Arts Portfolio (`arts.gov.au`) is a prime example, which transitioned from the top harmonic centrality cluster to the second. Originally part of the Department of Communication and Arts in 2018, encompassing national broadcasters and the postal service, it underwent a reorganization, subsequently integrating into the Department of Infrastructure, Transport, and Regional Development. Another case is the Bureau of Tourism Research (BTR) at btr.gov.au, which evolved into Tourism Research Australia (TRA) in 2006. Similarly, the Department of Communications and Arts (`dca.gov.au`) underwent a significant renaming and restructuring process. This scenario is typically characteristic of organizations, including government departments experiencing nomenclature changes, reorganizations, or abolition, such as dca.gov.au and government arts.gov.au. Additionally, it encompasses businesses that have ceased operations, like the Australian Motorcycle Museum (`australianmotorcyclemuseum.com.au`), or those undergoing name changes. This category also extends to entities operating in a seasonal or time-bound capacity, such as expos ((`townsvilleexpo.com.au`) or conferences such as (`sydneycubansalsacongress.com.au`), indicating a dynamic environment where organizational identities and structures are fluid and subject to change due to various internal and external factors.

## 5 DISCUSSION AND CONCLUSION

The health of a competitive economy depends on the existence of many independent firms that challenge each other. Traditionally, this has been ensured by two factors: the diminishing returns that limit the growth of monopolies, and the government regulation that prevents or breaks up excessive market power.

However, in the digital age, these factors are no longer sufficient. Network effects and platform strategies enable some firms to dominate entire industries and markets, without necessarily raising consumer prices. Examples of such firms are Uber, AirBnB and Google, which have gained near-monopoly positions in urban transport, accommodation and online advertising, respectively.

The phenomenon of Online Gravity, as described by Paul X. McCarthy in his book [14], explains how digital technologies have rebooted economics and created planet-like super-businesses that attract and retain most of the online attention and value. McCarthy also shows how network science can help understand and measure the evolution of diversity and dominance of companies in online activity [15], and how this can inform policy and strategy decisions. McCarthy's work also illustrates how in new industries online attention revealed through social media graph metrics can be linked to the growth in enterprise value of innovative companies like Tesla [15].

Table 4: Characteristics of organisations in whose domain stayed in the same cluster for six years from 2018 to 2023

|  | Stable Cluster 5 | Stable Cluster 4 | Stable Cluster 3 | Stable Cluster 2 | Stable Cluster 1 | Stable Cluster 0 |
|---|---|---|---|---|---|---|
| Characteristics | Established leading national and global consumer-facing brands, media, universities and Government. | Australian presence of global brands and national challenger brands. | Large suburban businesses in cities and metropolitan areas. | Local CBD businesses and niche speciality businesses. | Local businesses many in regional areas. | Local businesses many in country towns and remote areas. |
| Examples | Qantas, ASX, ATO, Australia Post, Woolworths, Coles, BoM, CSIRO, ABC, UNSW, University of Sydney | Loreal Australia, Ladbrokes Australia, DHL Australia, Foodworks. | RayWhite Drummoyne (real estate agency), Manly Cycles, Hobart Yachts, Perth Video | Vision blonde (hair salon), Viking Genetics (Cattle breeding) | Katie Smith Solicitor, Olivier Flowers (Central Coast), Northern Rivers Volkswagen (car dealer) | IGA Donnybrook (Donnybrook) |

To preserve economic diversity and consumer welfare, competition regulators need to adopt new approaches and criteria that go beyond price and market share. How to define and measure economic diversity is one of the challenges.

One influential advocate of this view is Lina Khan, the chair of the US Federal Trade Commission, who argued in her previous academic work [11] that antitrust law should focus on "the long-term interests of consumers [which] include product quality, variety and innovation — factors best promoted through both a robust competitive process and open markets."

Network science provides some useful tools and metrics to analyze the competitive dynamics, growth and influence of firms, especially in emerging sectors. One such metric is harmonic centrality, which measures the potential reach of a node in a network and is a measure of influence, trust and power within a network. In a competition setting, harmonic centrality may have applications in identifying firms that may have the most power and influence over other firms and thus can pose a potential threat to innovation and competition.

Harmonic centrality, a linkage-based metric, encapsulates the comprehensive web context of a domain along with its associated organization or brand. This study delved into deciphering domain-level and organization-level dynamics as illuminated by harmonic centrality. Our findings reveal that harmonic centrality is an intricate indicator, amalgamating data pertaining to the web graph structure at the website level, like domain size and inbound link quantity, with offline, non-structural elements such as organizational scale, reach, influence, and impact.

The detection of both structural and non-structural elements within harmonic centrality at a significant scale paves the way for intriguing future research prospects. One such avenue could involve inverting the inquiry to investigate whether harmonic centrality can predict organizational characteristics. Another promising direction might involve contrasting the harmonic centrality of entities within the same industry, correlating it with known metrics like scale, revenue, and industry influence. This approach aims to unravel additional, more subtle attributes potentially embedded within harmonic centrality, such as trustworthiness and authority.

Moreover, the consistency observed in the distributions of harmonic centrality across various nations suggests that our findings possess a degree of universality. This implication indicates that the insights derived from this study are not confined to the Australian web context but could be effectively generalized and applied to broader international settings. Consequently, this expands the applicability of our conclusions, underscoring the potential of harmonic centrality as a versatile tool in understanding and evaluating web-based and organizational phenomena on a global scale. Such research could significantly contribute to a deeper understanding of the interplay between web presence and organizational attributes, offering valuable perspectives for businesses and policymakers alike in the digital age.

## ACKNOWLEDGMENTS

The authors would like to acknowledge the generous advice and data provided for this study by Gary Brewer at BuiltWith. We would also like to thank the team at AuDA, especially Rosemary Sinclair, Bruce Tonkin, Sophie Mitchell and Michael Lewis for their vision in commissioning the related work that inspired this project and our wonderful collaborators Colin Griffith from CSIRO's Data61 and Claire McFarland from UTS who helped bring this foundation work to fruition. Author Paolo Boldi was supported in part by project SERICS (PE00000014) under the NRRP MUR program funded by the EU - NGEU.